\documentclass[aps,nofootinbib,superscriptaddress,twocolumn,floatfix,letterpaper,pra]{revtex4-1}
\usepackage{graphicx}
\usepackage{epstopdf}
\usepackage{amsmath,amsthm,amssymb,latexsym,amsfonts,times,mathrsfs,verbatim}
\usepackage{mathptmx}
\usepackage{braket}

\DeclareMathAlphabet{\mathcal}{OMS}{cmsy}{m}{n}

\renewcommand{\Re}{\ensuremath{\operatorname{Re}}}
\renewcommand{\Im}{\ensuremath{\operatorname{Im}}}

\begin{document}

\title{Phonon Quantum Nondemolition Measurements in Nonlinearly Coupled Optomechanical Cavities}

\author{B.D. Hauer}\email{bhauer@ualberta.ca}
\affiliation{Department of Physics, University of Alberta, Edmonton, Alberta, Canada T6G 2E9}
\author{A. Metelmann}
\affiliation{Dahlem Center for Complex Quantum Systems and Fachbereich Physik, Freie Universit\"{a}t Berlin, 14195 Berlin, Germany}
\author{J.P. Davis}\email{jdavis@ualberta.ca}
\affiliation{Department of Physics, University of Alberta, Edmonton, Alberta, Canada T6G 2E9}

\date{\today}

\begin{abstract}

In the field of cavity optomechanics, proposals for quantum nondemolition (QND) measurements of phonon number provide a promising avenue by which one can study the quantum nature of nanoscale mechanical resonators. Here, we investigate these QND measurements for an optomechanical system whereby quadratic coupling arises due to shared symmetries between a single optical resonance and a mechanical mode. We establish a relaxed limit on the amount of linear coupling that can exist in this type of system while still allowing for a QND measurement of Fock states. This new condition enables optomechanical QND measurements, which can be used to probe the decoherence of mesoscopic mechanical Fock states, providing an experimental testbed for quantum collapse theories.

\end{abstract}

\maketitle

\section{Introduction}
\label{intro}

The theory of quantum mechanics has excelled in describing a multitude of phenomena associated with microscopic systems. However, as a system scales to larger sizes, interaction with the surrounding environment causes its quantum mechanical state to decohere into the classical realm \cite{joos,schlosshauer}. Although there are a number of theories proposing mechanisms by which such decoherence could occur \cite{ghirardi,diosi,penrose,zurek}, this quantum-to-classical transition remains poorly understood, largely due to a lack of experimental systems that can be used to study these processes. To this end, a number of proposals have been put forward to use cavity optomechanics as an experimental platform to fill this void \cite{bose,marshall,romeroisart,chen}.

In cavity optomechanics, confined photonic degrees of freedom interact via radiation pressure with a mechanical resonator \cite{aspelmeyer}, allowing one to prepare the mechanical element into a variety of quantum states \cite{thompson, clerk2, gangat, vanner, borkje, riedinger}. Though experimental progress in quantum cavity optomechanics has been astounding \cite{teufel, chan, safavinaeini, safavinaeini2, palomaki, suh, weinstein, wollman, pirkkalainen, lecocq, lei, wilson, reed, hong, riedinger2, ockeleon-korppi}, a vital experiment still remains: the quantum nondemolition (QND) measurement of a mechanical resonator's phonon number \cite{thompson,gangat}. While QND measurements have been demonstrated for single particles \cite{bergquist,peil}, photons \cite{nogues,gleyzes}, spins \cite{neumann}, and superconducting qubits \cite{lupascu}, as well as for a single quadrature of a micromechanical resonator \cite{lecocq,lei}, measurements of the mechanical Fock states of a cavity optomechanical system would provide an engineerable platform to directly probe the decoherence of a mesoscopic quantum state. However, performing an experiment of this nature proves to be difficult, largely due to the fact that most optomechanical cavities couple linearly to the mechanical resonator's position \cite{aspelmeyer}. Such a scheme is unsuitable for QND measurements of the resonator's quantized energy, as it is subject to the Heisenberg uncertainty principle \cite{unruh,braginsky,braginsky2,clerk}. One must then turn to an optomechanical system where the optical mode is quadratically-coupled to the resonator's position, providing a method by which QND measurements of its phonon number can be performed \cite{thompson,gangat}.

\begin{figure}[t!]
\centerline{\includegraphics[width=\columnwidth]{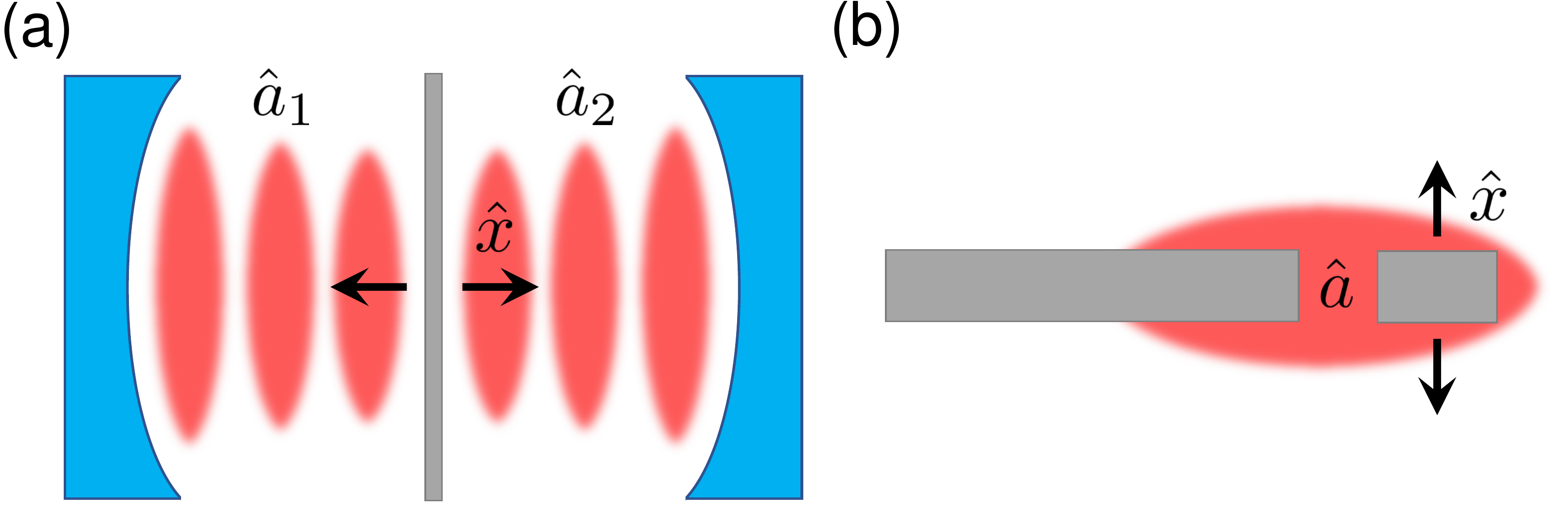}}
\caption{{\label{fig1}} Schematic of (a) a membrane-in-the-middle optomechanical system and (b) a mechanical element side-coupled to a whispering gallery mode optical cavity. In (a), quadratic coupling arises due to an avoided crossing between the two optical modes (see Appendex \ref{MIMWGM}), labeled by their creation operators $\hat{a}_1$ and $\hat{a}_2$. Meanwhile, in (b), the single optical mode denoted by $\hat{a}$ is coupled to the square of the mechanical motion via shared symmetries between the optics and mechanics (see Appendix \ref{nondegen}). The direction of the mechanical displacement, $\hat{x}$, is indicated by black arrows.}
\end{figure}

Experimental demonstration of quadratic coupling in optomechanical systems has been largely focussed on membrane-in-the-middle (MIM) systems (see Fig.~\ref{fig1}a), whereby a mechanical element -- which is typically a thin dielectric membrane \cite{thompson,sankey,jayich}, but can also be a cloud of cold atoms \cite{purdy} or a photonic crystal nanobeam \cite{paraiso,kalaee} -- is placed within an optical cavity. Inserting the ``membrane'' into the cavity causes its degenerate optical modes to hybridize into two new supermodes that exhibit an avoided level crossing \cite{jayich,paraiso}. By moving the membrane to an antinode of the optical mode, linear coupling of the membrane's motion is suppressed and quadratic coupling becomes dominant \cite{thompson,paraiso,jayich}. However, when optically driving one of these supermodes, parasitic linear coupling to the membrane's position emerges in the opposite mode, leading to an accelerated decoherence of the membrane's Fock state that can only be overcome in the single-photon strong-coupling regime \cite{miao,ludwig,yanay}. This stringent constraint has proven to be the most difficult obstacle to overcome when performing QND measurements of phonon number in a MIM optomechanical system \cite{paraiso}.

In this paper, we consider an optomechanical system that exhibits second-order coupling due to the shared symmetries between a mechanical resonance and a single optical mode \cite{kaviani}. Such a system could be physically realized as an out-of-plane flexural (or torsional) mode of a mechanical resonator side-coupled to a whispering gallery mode (WGM) of a microdisk \cite{doolin} (see Fig.~\ref{fig1}b) or the in-plane motion of a paddle located within a photonic crystal nanobeam \cite{kaviani}. Here, we analyze this quadratically-coupled optomechanical system in the context of resolving the thermally-induced jumps between the quantized energy eigenstates of the mechanical resonator. In doing so, we place a constraint on the relative strengths of the linear and quadratic optomechanical couplings of the system, demonstrating the parameter space for which a QND measurement of phonon number can be made. Furthermore, we show that in the case of a MIM system, this generalized limit is equivalent to the single-photon strong-coupling regime. The ability to perform these QND measurements will provide a direct probe of a mesoscopic quantum state, furthering our understanding of the mechanisms by which quantum systems transition into the classical regime. 

\section{Optomechanical Model}
\label{model}

We begin by modeling this optomechanical system as two coupled harmonic oscillators, with the Hamiltonian
\begin{equation}
\hat{H} = \hbar \omega_{\rm c} (\hat{x}) \hat{a}^\dag \hat{a} + \hbar \omega_{\rm m} \hat{b}^\dag\hat{b}.
\label{Ham1}
\end{equation}
Here, $\hat{a}$ and $\hat{b}$ ($\hat{a}^\dag$ and $\hat{b}^\dag$) are the annihilation (creation) operators of the optical cavity and the mechanical resonator, each with resonant angular frequencies of $\omega_{\rm c}(\hat{x})$ and $\omega_{\rm m}$, respectively. Coupling between the two oscillators arises due to the fact that the resonant frequency of the optical cavity is dependent on the position of the mechanics, $\hat{x} = x_{\rm zpf} \left( \hat{b} + \hat{b}^\dag \right)$, where $x_{\rm zpf} = \sqrt{\hbar/2 m \omega_{\rm m}}$ is the zero-point fluctuation amplitude of the mechanical oscillator, with $m$ being its effective mass \cite{hauerAOP1}. Expanding the position-dependence of the cavity frequency to second order, we obtain
\begin{equation}
\omega_{\rm c}(\hat{x}) = \omega_{\rm c} + G_1 \hat{x} + \frac{G_2}{2} \hat{x}^2,
\label{omcexp}
\end{equation}
where $\omega_{\rm c}$ is the unperturbed cavity frequency, along with the first- and second-order optomechanical coupling coefficients $G_1 = d\omega_{\rm c} /d\hat{x}$ and $G_2 = d^2\omega_{\rm c} / d\hat{x}^2$. Inputting Eq.~\eqref{omcexp}, as well as the expression for $\hat{x}$, into Eq.~\eqref{Ham1} results in
\begin{subequations}
\begin{align}
\label{Hamfulla}
\hat{H} &= \hat{H}_0 + \hat{H}', \\
\label{Hamfullb}
\hat{H}_0 &= \hbar \left[ \omega_{\rm c} + g_2 \left(\hat{b}^\dag \hat{b} + \frac{1}{2} \right)  \right] \hat{a}^\dag \hat{a} + \hbar \omega_{\rm m} \hat{b}^\dag \hat{b}, \\
\label{Hamfullc}
\hat{H}' &= \hbar g_1 \left( \hat{b} + \hat{b}^\dag \right) \hat{a}^\dag \hat{a} + \frac{\hbar g_2}{2} \left( \hat{b} \hat{b} + \hat{b}^\dag \hat{b}^\dag \right) \hat{a}^\dag \hat{a},
\end{align}
\end{subequations}
where we have introduced the single photon, single (two) phonon coupling rate $g_1 = G_1 x_{\rm zpf}$ ($g_2 = G_2 x_{\rm zpf}^2$). 

We have chosen to separate $\hat{H}$ into two sub-Hamiltonians, $\hat{H}_0$ and $\hat{H}'$, such that $\hat{H}_0$ commutes with the phonon number operator, $\hat{n} = \hat{b}^\dag \hat{b}$, while $\hat{H}'$ does not. In this way, $\hat{H}_0$ represents a QND measurement of the mechanical resonator's energy \cite{braginsky,braginsky2}, whereby changes in phonon number reflect a per phonon shift of $g_2$ in the optical cavity's resonant frequency. On the other hand, $[\hat{H}',\hat{n}] \ne 0$ such that $\hat{H}'$, which contains the interaction terms that evolve in time, acts to contaminate the QND measurement. This is a well-known fact for the first term in $\hat{H}'$, whereby linear coupling simultaneously probes the phase and energy of the mechanical resonator, preventing a QND Fock state measurement \cite{jayich,unruh}. In principle, one could completely eliminate this linear coupling by properly tuning the optical and mechanical symmetries of the system (see Appendix \ref{nondegen}). However, for any realistic optomechanical cavity, a non-zero amount of linear coupling will always creep into the system due to experimental inaccuracies \cite{thompson,doolin,kaviani}. Therefore, we seek to set a limit on the maximum allowable linear coupling that can exist in a quadratically-coupled optomechanical device that one wishes to use for a QND measurement of phonon number. Furthermore, we look to determine the regime for which the second term in $\hat{H}'$ can be safely ignored. To do this, we use a master equation approach to calculate the rates at which the linear and quadratic terms in $\hat{H}'$ act to cause the resonator to transition from a given Fock state. Comparing these rates to the mechanical resonator's intrinsic thermal decoherence rate and the optomechanical phonon state measurement rate, we ascertain the regime where QND measurements of phonon number can be achieved. 

\section{Mechanical Fock State Decoherence Rates}
\label{decrates}

To determine the relevant measurement-induced and thermal decoherence rates associated with the optomechanical QND measurement of mechanical phonon number considered in this work, we begin by assuming that the optical cavity is driven via a strong external drive with frequency $\omega_{\rm d}$. This displaces the optical cavity operators according to $\hat{a} = \bar{a} + \hat{d}$, where the $\bar{a}$ is the classical amplitude (taken to be real) and $\hat d$ denotes the cavity fluctuations. We then switch to a frame that rotates at the the optical cavity drive frequency by applying the unitary transform $\hat{U} = e^{i \omega_{\rm d} \hat{a}^\dag \hat{a}}$ to the Hamiltonian in Eq.~\eqref{Hamfulla} to obtain
\begin{align}
\hat{H} &= \hat{H}_{\rm c} + \hat{H}_{\rm m} + \hat{H}_{\rm om} + \hat{H}_\kappa + \hat{H}_\Gamma, \nonumber \\ 
\hat{H}_{\rm c} &= \hbar \Delta \hat{d}^{\dag} \hat{d}, \nonumber \\
\hat{H}_{\rm m} &= \hbar \omega_{\rm m} \hat{b}^{\dag} \hat{b}, \nonumber \\
\hat{H}_{\rm om} &=  \hbar \bar{a} \left[g_1 \left( \hat{b} + \hat{b}^{\dag} \right) + \frac{g_2}{2} \left(2 \hat{b}^{\dag} \hat{b} + \hat{b} \hat{b} + \hat{b}^{\dag} \hat{b}^{\dag} \right) \right] \left( \hat{d} + \hat{d}^{\dag} \right), \nonumber
\end{align}
where $\Delta = \omega_{\rm c} - \omega_{\rm d}$ is the detuning of the optical drive from the cavity's resonance frequency and we have neglected terms proportional to $\hat{d}^{\dag} \hat{d}$ in the interaction Hamiltonian $\hat{H}_{\rm om}$ \cite{aspelmeyer}. We note that we have now also included the coupling of the optical cavity and mechanical resonator to dissipative Markovian baths, which are described as usual \cite{clerk} and denoted by $\hat{H}_\kappa$ and $\hat{H}_\Gamma$, giving rise to a cavity decay rate of $\kappa$ and a mechanical daping rate of $\Gamma_{\rm m}$.

The dynamics of the system is then captured by the master equation, written in superoperator notation as \cite{carmichael}
\begin{equation}
 \frac{\partial}{\partial t} \hat{\rho} =  \left(  \mathcal{L}_{\rm c} +  \mathcal{L}_{\rm m} +  \mathcal{L}_{\rm om} \right) \hat{\rho},
\label{masterequation}
\end{equation}
with $\hat{\rho}$ being the total density matrix of the system (including both optical and mechanical components). Here, we have defined the superoperators
\begin{align}
\mathcal{L}_{\rm c} &= - \frac{i}{\hbar} [\hat{H}_{\rm c} , \bullet] + \frac{\kappa}{2} \mathcal D [\hat{d}] \bullet, \nonumber \\
\mathcal{L}_{\rm m} &= - \frac{i}{\hbar} [\hat{H}_{\rm m}  , \bullet ] + \frac{\Gamma_{\rm m}}{2} \left\{\left(\bar{n}_{\rm th} + 1 \right) \mathcal{D} [\hat{b}] + \bar{n}_{\rm th}  \mathcal{D} [\hat b^{\dag}]\right\} \bullet, \nonumber \\
\mathcal{L}_{\rm om} &= - \frac{i}{\hbar} [\hat{H}_{\rm om} , \bullet] , \nonumber \\
\mathcal{D}[\hat{o}] \bullet &= 2 \hat{o} \bullet \hat{o}^\dag - \hat{o}^\dag \hat{o} \bullet - \bullet \hat{o}^\dag \hat{o} = [\hat{o}, \bullet \hat{o}^\dag] + [\hat{o} \bullet, \hat{o}^\dag], \nonumber
\end{align}
where $\bullet$ is a placeholder for the operator the superoperator is acting upon, $\hat{o}$ is a generic ladder operator and we have introduced $\bar{n}_{\rm th} = \left( e^{\hbar \omega_{\rm m} / k_B T} - 1 \right)^{-1}$ as the average occupation number of the mechanical resonator's thermal bath at temperature $T$. Note that we have assumed a zero-temperature bath for the optical cavity.

We can now move into a new interaction picture with density matrix $\hat{\rho}' = e^{-(\mathcal{L}_{\rm c} + \mathcal{L}_{\rm m}) t } \hat{\rho}$ that evolves in time according to
\begin{equation}
\label{mastprime}
\frac{\partial \hat{\rho}'}{\partial t} = e^{- (\mathcal{L}_{\rm c} + \mathcal{L}_{\rm m}) t } \mathcal{L}_{\rm om} e^{ (\mathcal{L}_{\rm c} +  \mathcal{L}_{\rm m}) t } \hat{\rho}' \equiv \mathcal{L}_{\rm om}'(t) \hat{\rho}',
\end{equation}
where we have simply used the product rule and Eq.~\eqref{masterequation}. This master equation can be formally integrated to obtain the solution
\begin{equation}
 \hat{\rho}'(t) = \hat{\rho}'(0) + \int_{0}^{t} d\tau \mathcal{L}_{\rm om}'(\tau) \hat{\rho}'(\tau),
\end{equation}
which can be substituted back into Eq.~\eqref{mastprime} and, by additionally performing the trace over the cavity space, we arrive at the new master equation
\begin{equation}
\begin{split}
&\frac{\partial \hat{\rho}_{\rm m}'}{\partial t} \equiv \frac{\partial}{\partial t} {\rm Tr_c} \left\{ \hat{\rho}'(t) \right\} \\
&= {\rm Tr_c} \left\{ \mathcal{L}_{\rm om}'(t) \hat{\rho}'(0) \right\} + \int_{0}^{t} d\tau {\rm Tr_c} \left\{ \mathcal{L}_{\rm om}'(t)  \mathcal{L}_{\rm om}'(\tau)  \hat{\rho}'(\tau) \right\}.
\end{split}
\label{masttrace}
\end{equation}
where $\hat{\rho}'_{\rm m}$ is the density matrix of the mechanical resonator in the new interaction frame. Thus we have to calculate  
\begin{equation}
\mathcal{L}_{\rm om}' = - i \bar{a} \left[ \mathcal{A}(t) \mathcal{B}(t) - \mathcal{ A}^\dag(t) \mathcal{B}^\dag(t) \right],
\label{Lomprime}
\end{equation}
with
\begin{align}
\mathcal{A}(t) &= e^{-\mathcal{L}_{\rm c} t } \left( \hat{d} + \hat{d}^{\dag} \right)     \bullet e^{ \mathcal{L}_{\rm c} t }, \nonumber \\
\mathcal{B}(t) &= e^{-\mathcal{L}_{\rm m} t } \hat{B} \bullet e^{ \mathcal{L}_{\rm m} t }, \nonumber
\end{align}
where we have introduced the new mechanical operator $\hat{B} = g_1 \left( \hat{b} + \hat{b}^{\dag} \right) + \frac{g_2}{2} \left(2 \hat{b}^{\dag} \hat{b} + \hat{b} \hat{b} + \hat{b}^{\dag} \hat{b}^{\dag} \right)$.
Note that $\hat{B}$ and $\left( \hat{d} + \hat{d}^{\dag} \right)$ are Hermitian, but we have $ ( \hat{o} \bullet)^{\dag} = \bullet \hat{o}^{\dag}$, hence the appearance of 
$\mathcal{A}^\dag(t)$ and $\mathcal{B}^\dag(t)$ in Eq.~\eqref{Lomprime}. Finally, to evaluate the superoperator $\mathcal{L}_{\rm om}'$, we also need the dynamics of the cavity operator in this interaction picture: 
\begin{equation}
\mathcal{A}(t) = \hat{d} \bullet e^{- \left(i \Delta + \frac{\kappa}{2} \right) t} + \hat{d}^{\dag} \bullet e^{\left( i \Delta + \frac{\kappa}{2} \right) t} - \bullet \hat{d}^{\dag} e^{ i \Delta  t} \left( e^{\frac{\kappa}{2}t} - e^{-\frac{\kappa}{2}t}\right). \nonumber
\end{equation}

So far we have not made any approximations, the above treatment resembles the standard derivation for a master equation. In what follows, we adiabatically eliminate the cavity to obtain a reduced density matrix for the mechanics. This implies the assumption that the cavity photons adiabatically follow the phonon occupation, {\it i.e.}~that $\kappa \gg \Gamma_{\rm th}$ is fulfilled \cite{gangat}, where $\Gamma_{\rm th}$ is the thermal decoherence rate of the phonon state in question (see Eq.~\eqref{thermrate} below). Within this limit, we assume that the optical cavity and mechanical resonator are effectively uncorrelated at all times, so that the density matrix factorizes as $\hat{\rho} \equiv \hat{\rho}_{\rm m}  \otimes \hat{\rho}_{\rm c}$ \cite{martin}, where $\hat{\rho}_{\rm c}$ denotes the density matrix of the cavity mode. We also make a Born approximation and assume that the cavity mode fluctuations $\hat{d}$ are not affected by the dynamics of the mechanics, that is we set $\hat{\rho}_{\rm c}(t) \approx \hat{\rho}_{\rm c}(0)$. This means that the total density matrix remains a product of the initial cavity density matrix and the mechanical density matrix, {\it i.e.}~$\hat{\rho}'(t) \approx \hat{\rho}_{\rm m}'(t) \otimes \hat{\rho}_{\rm c}(0) \equiv \hat{\rho}_{\rm m}'(t) \otimes |0\rangle \langle 0|$ (for the cavity being in the vacuum state in this displaced frame). Under this assumption the first term in Eq.~\eqref{masttrace} vanishes and  with
\begin{align}
{\rm Tr_c} \left\{ \mathcal{A}(t) \mathcal{A}(\tau) \; |0\rangle \langle 0| \right\} &= {\rm Tr_c} \left\{ \mathcal{A}^\dag(t) \mathcal{A}(\tau) \; |0\rangle \langle 0| \right\} \nonumber \\ 
&= e^{- \left(i \Delta + \frac{\kappa}{2} \right) (t-\tau) } , \nonumber \\ 
{\rm Tr_c} \left\{ \mathcal{A}^\dag(t) \mathcal{A}^\dag(\tau) \; |0\rangle \langle 0|  \right\} &= {\rm Tr_c} \left\{ \mathcal{A}(t) \mathcal{A}^\dag(\tau) \; |0\rangle \langle 0|  \right\} \nonumber \\ 
&= e^{+\left(i \Delta - \frac{\kappa}{2} \right) (t-\tau) }, \nonumber
\end{align}
we can evaluate the second term as
\begin{equation}
\label{secterm}
\begin{split}
&{\rm Tr_c} \left\{ \mathcal{L}_{\rm om}'(t)  \mathcal{L}_{\rm om}'(\tau) \hat{\rho}'(\tau)\right\} = \\ 
&- \bar{N} \left[ \left\{ \mathcal{B}(t) \mathcal{B}(\tau) - \mathcal{B}^\dag(t)   \mathcal{B}(\tau) \right\} e^{-\left(i \Delta + \frac{\kappa}{2} \right) (t-\tau) } + {\rm h.c.} \right] \hat{\rho}_{\rm m}'(\tau).
\end{split}
\end{equation}
Here 
\begin{equation}
\bar{N} = \frac{\kappa_{\rm e}}{\Delta^2 + \left( \kappa / 2 \right)^2} \frac{P}{\hbar \omega_{\rm d}},
\label{Nave}
\end{equation}
is the average intracavity photon number, with $\kappa_{\rm e}$ and $P$ being the optical cavity's external decay rate and input power. Using the expression in Eq.~\eqref{secterm}, the master equation yields (with change of variables $t' = t -\tau$)
\begin{equation}
\begin{split}  
\frac{\partial \hat{\rho}_{\rm m}'}{\partial t} = - \bar{N}& \int_{0}^{t} dt' \; \hat{\rho}_{\rm m}'(t-t') \Big[ \big\{ \mathcal{B}(t) \mathcal{B}(t-t') \\ 
&- \mathcal{B}^\dag(t) \mathcal{B}(t-t') \big\} e^{- \left(i \Delta + \frac{ \kappa}{2} \right) t'} + {\rm h.c.} \Big],
\end{split}
\end{equation}
which we can transform back to the initial frame knowing that $\hat{\rho}_{\rm m}(t)  = e^{\mathcal{L}_{\rm m}t } \hat{\rho}_{\rm m}'(t)$ and move into an interaction picture with respect to the free mechanical Hamiltonian. This gives us 
\begin{equation}
\begin{split}
&\frac{\partial \hat{\rho}_{\rm m}}{\partial t} = - \bar{N} \int_{0}^{t} dt' \; \hat{\rho}_{\rm m} (t-t') \bigg( \bigg\{g_1^2 \Big[ \left\{ \hat{b} \hat{b}^\dag \bullet - \hat{b}^\dag \bullet   \hat{b} \right\} e^{- i \omega_{\rm m} t'} \\ 
&+ \left\{ \hat{b}^\dag \hat{b} \bullet - \hat{b} \bullet \hat{b}^\dag \right\} e^{+ i \omega_{\rm m} t'} \Big] + g_2^2 \left[\hat{b}^{\dag} \hat{b} \hat{b}^{\dag} \hat{b} \bullet - \hat{b}^{\dag} \hat{b} \bullet \hat{b}^{\dag} \hat{b} \right] \\ &+ \frac{g_2^2}{4} \Big[ \left\{\hat{b} \hat{b} \hat{b}^\dag  \hat{b}^\dag \bullet - \hat{b}^\dag \hat{b}^\dag \bullet \hat{b} \hat{b} \right\} e^{- i 2 \omega_{\rm m} t'} \\ 
&+ \left\{\hat{b}^\dag \hat{b}^\dag \hat{b} \hat{b} \bullet - \hat{b} \hat{b} \bullet \hat{b}^\dag \hat{b}^\dag \right\} e^{+ i 2 \omega_{\rm m}  t'} \Big] \bigg\} e^{- \left(i \Delta + \frac{\kappa}{2} \right) t'} + {\rm h.c.} \bigg) \\
&+ \frac{\Gamma_{\rm m}}{2} \left\{ \left( \bar{n}_{\rm th} + 1 \right) \mathcal{D}[\hat{b}] + \bar{n}_{\rm th} \mathcal{D}[\hat{b}^\dag] \right\} \hat{\rho}_{\rm m}(t), \nonumber
\end{split}
\end{equation}
where we have also used the rotating wave approximation \cite{aspelmeyer}. In the next step we apply a Markov approximation and solve the integrals for $t \rightarrow \infty$, obtaining
\begin{equation}
\begin{split} 
&\frac{\partial \hat{\rho}_{\rm m}}{\partial t} = -\frac{i \bar{N}}{\hbar} [ \hat{H}_{\rm r}, \hat{\rho}_{\rm m} (t) ] + \bar{N} \; \bigg(g_1^2 \Re \left\{ \chi_{\rm c}(\omega_{\rm m}) \right\} \mathcal{D} [\hat{b}^\dag] \\ 
&+ g_1^2 \Re \left\{ \chi_{\rm c}(- \omega_{\rm m}) \right\} \mathcal{D}[\hat{b}] + g_2^2 \Re \left\{ \chi_{\rm c}(0) \right\} \mathcal{D} [\hat{b}^\dag \hat{b}] \\
&+ \frac{g_2^2}{4} \Re \left\{ \chi_{\rm c}( 2\omega_{\rm m}) \right\} \mathcal{D}[\hat{b}^\dag \hat{b}^\dag] + \frac{g_2^2}{4} \Re \left\{ \chi_{\rm c}(-2\omega_{\rm m}) \right\} \mathcal{D}[\hat{b} \hat{b}] \bigg) \hat{\rho}_{\rm m} (t) \\ 
&+ \frac{\Gamma_{\rm m}}{2} \left\{ \left(\bar{n}_{\rm th} + 1 \right) \mathcal{D}[\hat{b}] +  \bar{n}_{\rm th} \mathcal{D}[\hat{b}^\dag] \right\} \hat{\rho}_{\rm m} (t).
\end{split}
\label{rhomfin}
\end{equation} 
Here, $\chi_{\rm c}(\omega) = [i ( \Delta + \omega  ) + \kappa/2]^{-1}$ is the susceptibility of the optical cavity, with real and imaginary parts
\begin{align}
\Re \left\{ \chi_{\rm c}(\omega) \right\} &= \frac{ \kappa / 2}{(\omega + \Delta)^2 + (\kappa / 2)^2}, \nonumber \\
 \hspace{0.5cm}
\Im \left\{ \chi_{\rm c}(\omega) \right\} &= -\frac{(\omega + \Delta)}{(\omega + \Delta)^2 + (\kappa / 2)^2}. \nonumber
\end{align}
Note that $\Re \left\{ \chi_{\rm c}(\omega) \right\} = S_{NN}(-\omega) / 2 \bar{N}$, with 
\begin{equation}
S_{NN}(\omega) =  \frac{\bar{N} \kappa}{\left( \omega - \Delta \right)^2 + \left( \kappa / 2 \right)^2},
\label{SNNdef}
\end{equation}
being the photon number spectral density \cite{thompson,marquardt,clerk,aspelmeyer}. 

The coherent dynamics in Eq.~\eqref{rhomfin} are described by the Hamiltonian
\begin{equation}
\begin{split}
&\hat{H}_{\rm r} = \hbar g_1^2 \left( \Im \left\{ \chi_{ \rm c}(\omega_{\rm m}) \right\}  + \Im  \left\{ \chi_{\rm c}(-\omega_{\rm m}) \right\} \right)  \hat{b}^\dag \hat{b} \\ 
&+ \hbar  g_2^2 \Im \left\{  \chi_{\rm c}(0) \right\} \hat{b}^\dag  \hat{b} \hat{b}^\dag   \hat{b} + \frac{\hbar g_2^2}{4} \Big( \Im \left\{ \chi_{\rm c}(2 \omega_{\rm m}) \right\} \hat{b} \hat{b} \hat{b}^\dag \hat{b}^\dag \\
&+ \Im \left\{ \chi_{\rm c}(-2 \omega_{\rm m}) \right\} \hat{b}^\dag \hat{b}^\dag \hat{b} \hat{b} \Big), \nonumber
\end{split}
\end{equation}
where the first term describes a shift induced by the linear coupling, while the second and third terms are of the Kerr-type (Lamb shifts). Note that this photon-mediated coherent interaction does not affect the phonon occupation, {\it i.e.}~$[\hat{H}_{\rm r},\hat{b}^{\dag} \hat{b}] =  0$. Furthermore, for zero detuning ($\Delta = 0$) the above Hamiltonian simplifies to $\hat{H}_{\rm r} = \hbar g_2^2 \Im \left\{ \chi_{\rm c}( 2 \omega_{\rm m}) \right\} \left( \hat{b}^{\dag} \hat{b} + \frac{1}{2} \right)$, resulting in a static shift of the mechanical frequency. Eq.~\eqref{rhomfin} describes the dynamics of the mechanical mode under influence of the QND measurement, as well as it's contamination due to the linear coupling, counter-rotating second-order terms and the influence of the thermal environment. The mechanical occupation is not affected by the pure QND measurement as expected, {\it i.e.}~terms associated with the susceptibility on resonance, $\chi_{\rm c}(0)$, in Eq.~\eqref{rhomfin} do not change the occupation.

We now use the fact that we can determine the probability, $p_n$, of being in the $n$th Fock state by taking the inner product of the density matrix using the number state basis vectors, that is $p_n(t) = \bra{n} \hat{\rho}_{\rm m}(t) \ket{n}$. If we assume that we are in the initially in the $n$th Fock state, such that $p_n(0) = 1$, then the total rate at which the mechanical resonator decoheres from this pure state can be found using Eq.~\eqref{rhomfin}. This results in
\begin{equation}
\left| \dot{p_n} \right| = \left| \bra{n} \dot{\rho}_{\rm m} \ket{n} \right| = \Gamma_{n+1} + \Gamma_{n-1} + \Gamma_{n+2} + \Gamma_{n-2} + \Gamma_{\rm th}. \nonumber
\end{equation}
where
\begin{align}
\label{Gamn+1}
\Gamma_{n+1} &= (n+1) g_1^2 S_{NN}(-\omega_{\rm m}), \\
\label{Gamn-1}
\Gamma_{n-1} &= n g_1^2 S_{NN}(\omega_{\rm m}), \\
\label{Gamn+2}
\Gamma_{n+2} &= (n+1)(n+2) \frac{g_2^2}{4} S_{NN}(-2 \omega_{\rm m}), \\
\label{Gamn-2}
\Gamma_{n-2} &= n (n-1) \frac{g_2^2}{4} S_{NN}(2 \omega_{\rm m}),
\end{align}
are the rates at which the phonon state of the mechanical resonator decoheres due to measurement-induced jumps from $n \rightarrow n \pm 1$ and $n \rightarrow n \pm 2$, while
\begin{equation}
\Gamma_{\rm th} = \Gamma_{\rm m} \left[ \left(\bar{n}_{\rm th} + 1 \right) n + \bar{n}_{\rm th} \left( n+ 1 \right) \right],
\label{thermrate}
\end{equation}
is the rate associated the thermal decoherence of the mechanical resonator's $n$th Fock state due to coupling with its dissipative bath \cite{gardiner,santamore}. We note that this thermal decoherence rate can be decreased by reducing the thermal occupation of its bath as low as possible. Furthermore, as can be seen in Fig.~\ref{fig2}, each of the decoherence rates given Eqs.~\eqref{Gamn+1}-\eqref{thermrate} decrease as we move to lower Fock states, taking on their minimum values for a mechanical resonator in its ground state.

\section{QND Measurement Conditions}
\label{qndcond}

In order to temporally resolve jumps between the mechanical resonator's phonon number states, one must measure the system faster than it decoheres. It was shown in Ref.~\cite{gangat} that for QND measurements of mechanical energy, the quadratic optomechanical measurement will collapse the system into a phononic number state at a rate given by
\begin{equation}
\Gamma_{\rm meas} = \bar{C}_2 \Gamma_{\rm m},
\label{measrate}
\end{equation}
where $\bar{C}_2 = \bar{N} C_2$ is the second-order, cavity-enhanced cooperativity of the system, given in terms of the second-order, single-photon cooperativity, $C_2 = 4 g_2^2 / \kappa \Gamma_{\rm m}$ \cite{aspelmeyer}. Comparing this measurement rate to the decoherence rates found in Eqs.~\eqref{Gamn+1}-\eqref{thermrate}, one finds the following hierarchy required to perform optomechanical QND measurements of mechanical energy quantization:
\begin{equation}
\Gamma_{\rm meas} \gg \Gamma_{\rm th} \gg \Gamma_{n \pm 1}, \Gamma_{n \pm 2}.
\label{ratehier}
\end{equation}
The right hand side of Eq.~\eqref{ratehier} ensures that thermal transitions dominate over optically-induced phonon jumps, leading to the ``linear-coupling condition'', $\Gamma_{\rm th} \gg \Gamma_{n \pm 1}$, and the ``quadratic-coupling condition'', $\Gamma_{\rm th} \gg  \Gamma_{n \pm 2}$. In this situation, one would expect a phonon distribution resembling a thermal state \cite{gangat}, exhibiting Bose-Einstein statistics with an average phonon occupation of $\braket{n} = \bar{n}_{\rm th}$. However, if one were to enter a regime where $\Gamma_{\rm th} \lesssim \Gamma_{n \pm 1},\Gamma_{n \pm 2}$, phonon trajectories would be dominated by optomechanically-induced jumps, leading to far more complex phonon statistics.

\begin{figure}[t!]
\centerline{\includegraphics[width=\columnwidth]{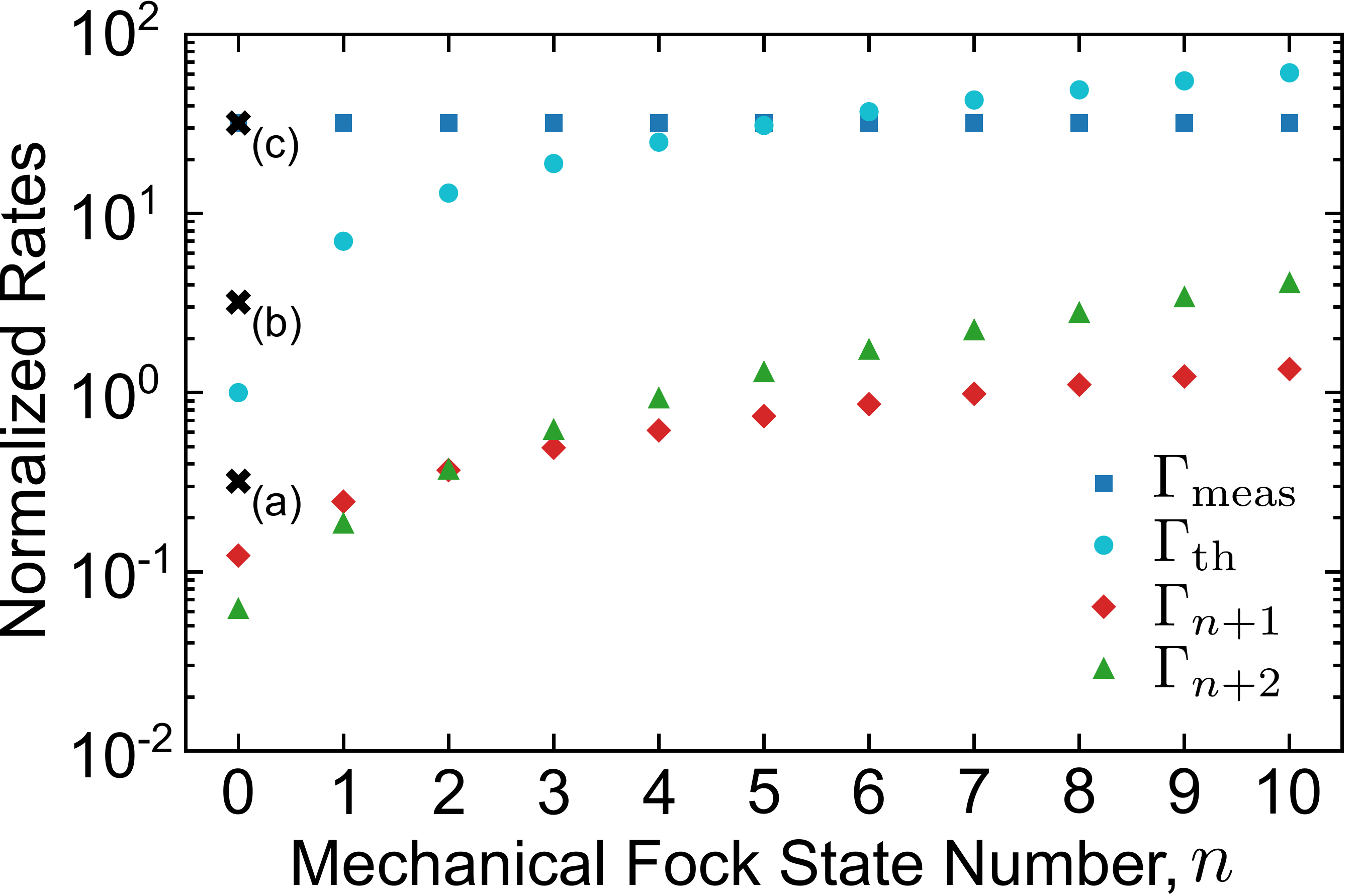}}
\caption{{\label{fig2}} Plot of the measurement rate, $\Gamma_{\rm meas}$, and thermal decoherence rate, $\Gamma_{\rm th}$, as well as the first- and second-order measurement-induced transition rates, $\Gamma_{n+1}$ and $\Gamma_{n+2}$, normalized to the thermal decoherence rate of the ground state, $\Gamma_{\rm th}^0 = \bar{n}_{\rm th} \Gamma_{\rm m}$, for the first ten Fock states of an optomechanical system. The system parameters are: $\omega_{\rm m} / 2 \pi =$ 2 GHz, $\Gamma_{\rm m} / 2 \pi =$ 1 kHz ($Q_{\rm m} =$ 2 $\times$10$^6$), $\bar{n}_{\rm th}$ = 0.25 ($T \approx$ 60 mK), $\Delta = 0$, $\kappa / 2 \pi =$ 500 MHz, $\bar{N} =$ 100, $g_1 / 2 \pi =$ 50 kHz, $g_2 / 2 \pi =$ 100 kHz. Here, $n_{\rm max} \approx$ 5 such that the first five mechanical Fock states can be monitored continuously using the QND measurement discussed in the main text. The black crosses indicate the measurement rate values used for the trajectory simulations in Fig.~\ref{fig3}.}
\end{figure} 

\begin{figure}[t!]
\centerline{\includegraphics[width=\columnwidth]{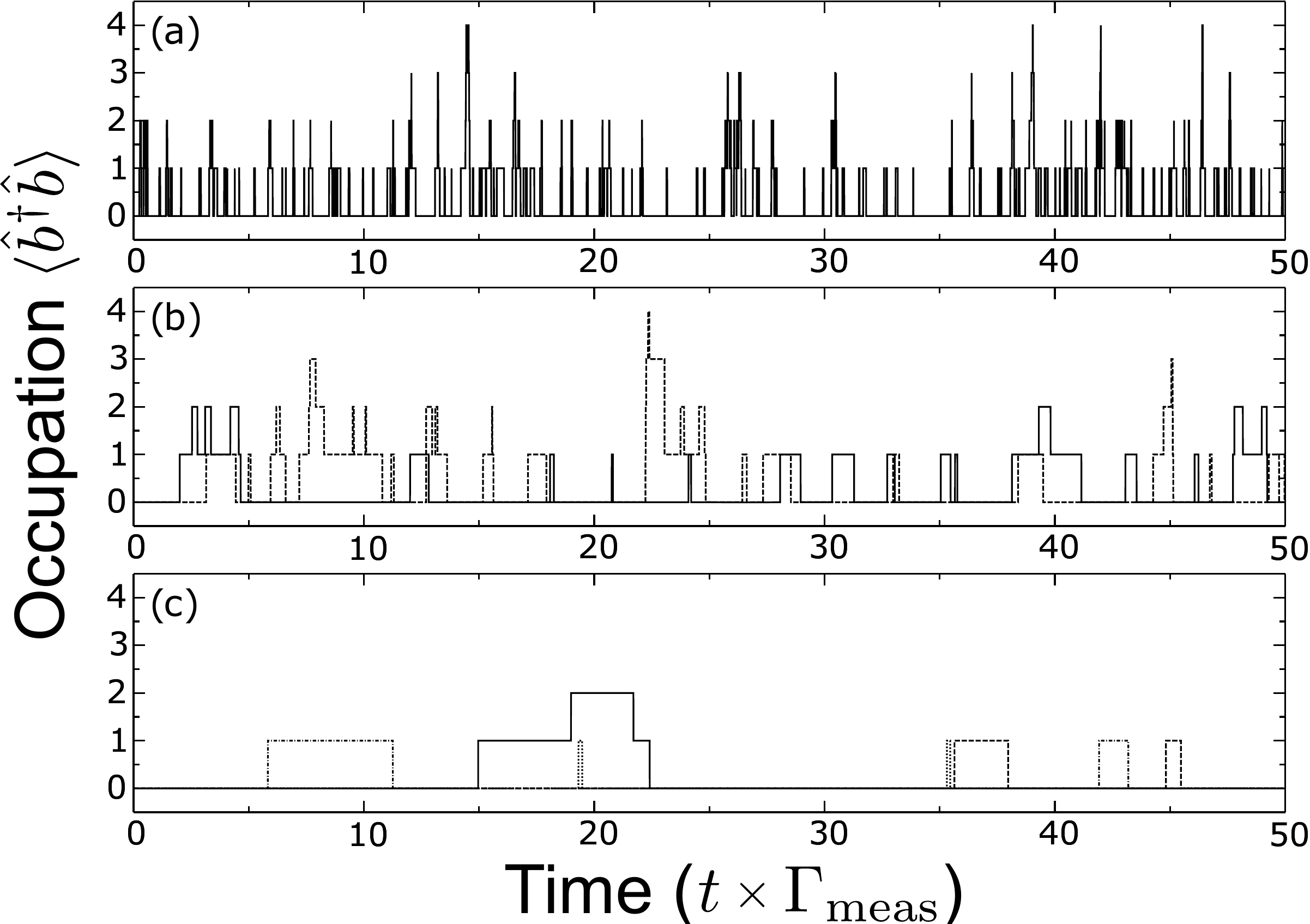}}
\caption{{\label{fig3}} 
Monte Carlo simulations of the mechanical occupation dynamics using the master eqation in Eq.~\eqref{rhomfin} for three measurement rates, $\Gamma_{\rm meas}/\Gamma_{\rm th} =$ (a) 0.32, (b) 3.2, and (c) 32. The remaining parameters are those found in Fig.~\ref{fig2}. In each plot we have included (a) a single trajectory, (b) two trajectories, and (c) four trajectories. Graph (a) depicts the case where thermally-induced jumps dominate. Here, a QND measurement is not possible and the final detected signal would only give information about the average phonon number. In contrast, if the inverse measurement rate is the smallest time scale in the system, as seen in (c), the QND read-out of the occupation can occur fast enough to resolve quantum jumps in phonon number. We note that in an actual experiment, the phonon trajectory would be inferred from a physical observable, such as the current from a homodyne detector \cite{gangat}.}
\end{figure}

Also included in Eq.~\eqref{ratehier} is the ``fast-measurement condition'' $\Gamma_{\rm meas} \gg \Gamma_{\rm th}$ \cite{gangat}, which tells us that one must be able to measure the phonon state of the resonator before it thermally decoheres in order to resolve quantized mechanical energy jumps \cite{martin,jayich,thompson,yanay}. To confirm this condition, we have performed Monte Carlo simulations of mechanical phonon trajectories according to the master equation in Eq.~\eqref{rhomfin} \cite{johansson}. As can be seen in Fig.~\ref{fig3}, one can enter into a regime where the optomechanical measurement rate is fast enough to allow for observation of quantum jumps in mechanical phonon number. We note that the fast-measurement condition can also be used determine the largest Fock state number that can be continuously monitored using this QND scheme as
\begin{equation}
n_{\rm max} = \frac{\bar{C}_2 - \bar{n}_{\rm th}}{2 \bar{n}_{\rm th} + 1}.
\label{nmax}
\end{equation}

The above analysis is valid for any arbitrary Fock state of the mechanical resonator. However, the minimum requirement necessary to perform a QND measurement of the mechanical oscillator's energy will occur when the system is in its ground state, as $\Gamma_{\rm th}$, $\Gamma_{n \pm 1}$ and $\Gamma_{n \pm 2}$ are all indeed minimized for $n=0$ (see Fig.~\ref{fig2}). In this situation, we are no longer concerned with rates corresponding to a reduction in phonon number ({\it i.e.}~$\Gamma_{n - 1}$, $\Gamma_{n - 2}$), as the mechanical ground state is unable to emit phononic energy. Furthermore, in order to experimentally resolve shifts in the optical cavity resonance frequency due to the creation/annihilation of a single phonon, one often turns to a phase sensitive transduction scheme, such as optical homodyne detection \cite{schliesser}, where signal is maximized at $\Delta =0$. Finally, we wish to operate in the sideband-resolved regime ($\omega_{\rm m} \gg \kappa$), which as we shall see below is necessary for the QND measurements of the mechanical ground state. Under these circumstances, the rate hierarchy of Eq.~\eqref{ratehier} becomes
\begin{equation}
\Gamma_{\rm meas} \gg \Gamma_{\rm th}^0 \gg \Gamma_1, \Gamma_2,
\label{ratehierGS}
\end{equation}
where $\Gamma_{\rm th}^0 = \bar{n}_{\rm th} \Gamma_{\rm m}$ is the rate at which the ground state of the mechanical resonator thermally decoheres, found by taking $n=0$ in Eq.~\eqref{thermrate}. Furthermore, $\Gamma_1$ and $\Gamma_2$ are the measurement-induced rates associated with transitions to the first and second excited states from this ground state at $\Delta = 0$ and are given by
\begin{align}
\label{Gam1}
\Gamma_1 &= \frac{\bar{N} g_1^2 \kappa}{\omega_{\rm m}^2} = \frac{\bar{C}_1 \Gamma_{\rm m} \kappa^2}{4 \omega_{\rm m}^2}, \\
\label{Gam2}
\Gamma_2 &= \frac{\bar{N} g_2^2 \kappa}{8 \omega_{\rm m}^2} = \frac{\bar{C}_2 \Gamma_{\rm m} \kappa^2}{32 \omega_{\rm m}^2},
\end{align}
where similar to before we have introduced the first-order, cavity-enhanced cooperativity, $\bar{C}_1 = \bar{N} C_1$, in terms of the corresponding first-order, single-photon cooperativity, $C_1 = 4 g_1^2 / \kappa \Gamma_{\rm m}$. Using these two expressions for $\Gamma_1$ and $\Gamma_2$, the limits in Eq.~\eqref{ratehierGS} can also be cast in terms of the first- and second-order quantum cooperativities, $\mathcal{C}_1 = \bar{C}_1 / \bar{n}_{\rm th}$ and $\mathcal{C}_2 = \bar{C}_2 / \bar{n}_{\rm th}$ \cite{aspelmeyer}, as
\begin{equation}
\mathcal{C}_2 \gg 1 \gg \mathcal{C}_1 \left( \frac{\kappa^2}{4 \omega_{\rm m}^2} \right), \mathcal{C}_2 \left( \frac{\kappa^2}{32 \omega_{\rm m}^2} \right).
\label{ratehierGScoop}
\end{equation}

We now look to understand the fundamental limits associated with this type of QND measurement. First, by ensuring that $\Gamma_{\rm meas} \gg \Gamma_2$ -- that is to say we can measure the phononic state of the system before an optically-induced transition to the second excited state occurs -- we arrive at the condition $32 \omega_{\rm m}^2 \gg \kappa^2$, which will certainly be satisfied for a sideband-resolved optomechanical system. Physically, this limit can be interpreted as cavity photons have a lifetime much longer than the mechanical period, such that they sample the mechanical motion over many cycles. This effectively averages out rapidly oscillating, transition-inducing second-order terms in the Hamiltonian in favor of terms that are constant in time.

Furthermore, the requirement that $\Gamma_{\rm meas} \gg \Gamma_1$ allows us to set the following limit on the linear coupling with respect to the quadratic coupling
\begin{equation}
g_2 \gg g_1 \frac{\kappa}{2 \omega_{\rm m}}.
\label{linlim}
\end{equation}
By satisfying this inequality, optomechanical QND measurements of phonon number can be performed. Upon further inspection of Eq.~\eqref{linlim}, we see that the sideband resolution condition discussed in the previous paragraph aids in suppressing the detrimental effect of linear coupling. A slightly more subtle observation is that since $g_2$ is proportional to $x_{\rm zpf}^2$ while $g_1$ is linear in $x_{\rm zpf}$, larger zero-point fluctuation amplitudes (corresponding to smaller effective masses and mechanical resonance frequencies) act to further relax the condition of small linear coupling. We note that in order to satisfy this limit, the relative strengths of $g_1$ and $g_2$ (or equivalently $G_1$ and $G_2$) must be able to be tuned independently. This can be done for the system considered here by utilizing the symmetry between the optics and mechanics (see Appendix \ref{nondegen}). However, this is not the case for quadratic coupling resulting from the hybridization of two nearly degenerate optical modes, as is found in MIM systems, where $g_2 = g_1^2 / 2 \nu$ with $\nu$ being the coupling rate between the two optical modes (see Appendix \ref{MIMWGM}). In fact, one can put this relation into Eq.~\eqref{linlim} to recover the single-photon strong-coupling requirement $g_1 \gg \kappa$, where we have assumed $2\nu \gg \omega_{\rm m}$ as is regularly done with MIM systems \cite{miao,ludwig,yanay,paraiso}. Therefore, Eq.~\eqref{linlim} provides a more general, less stringent condition for QND measurements of phonon number using a quadratically-coupled optomechanical cavity.

As a final note, we point out that even if the above conditions are met, one must still satisfy the ground state linear- and quadratic-coupling conditions, {\it i.e.}~$\Gamma_{\rm th}^0 \gg \Gamma_1, \Gamma_2$. Therefore, the greater the sideband resolution and linear coupling suppression in a given system, the larger the difference between $\Gamma_{\rm meas}$ and both measurement-induced transition rates, $\Gamma_1$ and $\Gamma_2$, producing a larger range of ground state thermal decoherence rates that will satisfy Eq.~\eqref{ratehierGS}. Furthermore, we emphasize the importance of low thermal bath occupation. Even if one can cool the mechanical mode to near its ground state using active feedback cooling techniques \cite{rocheleau,teufel,chan,riviere}, it is not possible to reduce the thermal decoherence rate of a given Fock state below $\Gamma_{\rm th}^0 = \bar{n}_{\rm th} \Gamma_{\rm m}$. Therefore, passive cooling of an optomechanical system using a refrigeration system \cite{riviere2,meenehan,macdonald,macdonald2} will likely be necessary to facilitate these types of continuous QND measurements.

\section{Conclusion}
\label{conc}

In this paper, we have investigated the limits involved with performing QND measurements of the quantized mechanical Fock states in an optomechanical cavity where quadratic coupling arises due to shared symmetries between a single optical and mechanical mode. By imposing the requirement that one measures the phononic state of the system faster than it thermally decoheres or transitions to another Fock state via the optomechanical interaction itself, it was shown that the single-photon strong-coupling condition associated with MIM systems can be circumvented. Instead a new, less stringent limit on the strength of the linear coupling was imposed, along with optomechanical sideband resolution. With these conditions satisfied, such an optomechanical system can be used to perform quantum jump spectroscopy \cite{peil} on the thermally-induced transitions between mechanical quanta. One could also consider using this type of QND measurement to freeze the resonator into a given Fock state, prolonging its coherence time via the quantum Zeno effect \cite{misra}, as has been demonstrated for trapped ions \cite{itano} and cold atoms \cite{fischer,patil}. Such an effect could be useful for a number of optomechanical quantum information protocols \cite{wang,stannigel}, where long coherence times are beneficial for applications such as quantum memories \cite{riedinger,reed} and transducers \cite{hill,bochmann,andrews,palomaki2}. Furthermore, the ability to observe and manipulate the decoherence of these mesoscopic quantum mechanical states would provide a long sought-after experimental platform to aid in the understanding of the elusive quantum-to-classical transition.

Finally, we note that the type of QND measurements considered in this work could also be applied to electromechanical systems, as was considered recently in Ref. \cite{dellantonio}, where the authors arrive at results similar to Eq.~\eqref{linlim} of this paper.

\begin{acknowledgments}

The authors would like to thank Saeed Khan for bringing to our attention the master equation treatment that was used to calculate the decoherence rates in Sec.~\ref{decrates}.

This work was supported by the University of Alberta, Faculty of Science; the Natural Sciences and Engineering Research Council, Canada (Grants Nos. RGPIN-2016-04523, DAS492947-2016, and STPGP 493807 - 16); and the Canada Foundation for Innovation.  B.D.H. acknowledges support from the Killam Trusts. A.M. acknowledges support from by the Deutsche Forschungsgemeinschaft through the Emmy Noether program (Grant No. ME 4863/1-1).

\end{acknowledgments}

\begin{appendix}

\section{Comparison of Membrane-in-the-Middle and Whispering Gallery Mode Optomechanical Systems}
\label{MIMWGM}

\subsection{Two Mode Optomechanical Hamiltonian}
\label{twomode}

Here, we consider a mechanical resonator, with resonant angular frequency $\omega_{\rm m}$, position operator $\hat{x}$ and phononic annihilation (creation) operators $\hat{b}$ ($\hat{b}^\dag$), that is simultaneously coupled to two optical modes, each of which are characterized by the annihilation (creation) operators $\hat{a}^{\phantom \dag}_1$ ($\hat{a}^\dag_1$) and $\hat{a}^{\phantom \dag}_2$ ($\hat{a}^\dag_2$), as well as the position-dependent angular frequencies $\omega_1(\hat{x})$ and $\omega_2(\hat{x})$. We also allow these two optical modes to be coupled to each other at a rate $\nu$, which physically manifests itself as a photon tunneling rate between the optical modes to the left and right of the membrane in membrane-in-the-middle (MIM) systems \cite{jayich,paraiso,ludwig} or a backscattering rate between clockwise and counterclockwise propagating modes in whispering gallery mode (WGM) optomechanics \cite{kippenberg}. The Hamiltonian for such an optomechanical system (ignoring the ground state energies, drive terms and interaction with the environment) will be given by
\begin{equation}
\begin{split}
&\hat{H} = \hbar \omega_1(\hat{x}) \hat{a}_1^\dag \hat{a}^{\phantom \dagger}_1 + \hbar \omega_2 (\hat{x}) \hat{a}_2^\dag \hat{a}^{\phantom \dagger}_2 + \hbar \omega_{\rm m} \hat{b}^\dag \hat{b} \\ 
&+ \hbar \nu \left( \hat{a}_1^\dag \hat{a}_2^{\phantom \dagger} + \hat{a}_2^\dag \hat{a}_1^{\phantom \dagger} \right),
\label{Htwoopt}
\end{split}
\end{equation}
where the last term describes the interaction between the two optical modes, with a photon being annihilated in one while simultaneously created in the other. As was done in Sec.~\ref{model}, we expand each $i$th optical frequency to second order in mechanical position as 
\begin{equation}
\omega_i(\hat{x}) = \omega_i + G_1^{(a_i)} \hat{x} + \frac{G_2^{(a_i)}}{2} \hat{x}^2,
\label{omtwoopt}
\end{equation}
where again we have the unperturbed optical frequency $\omega_i$, as well as the first- and second-order optomechanical coupling coefficients, $G_1^{(a_i)}$ and $G_2^{(a_i)}$, with the superscript $(a_i)$ allowing for one to identify the coupling coefficient associated with each optical mode. Inputting these expressions into Eq.~\eqref{Htwoopt}, we obtain the interaction Hamiltonian for the system
\begin{equation}
\begin{split}
&\hat{H} = \hbar \omega_1 \hat{a}_1^\dag \hat{a}^{\phantom \dagger}_1 + \hbar \omega_2 \hat{a}_2^\dag \hat{a}^{\phantom \dagger}_2 + \hbar \omega_{\rm m} \hat{b}^\dag \hat{b} \\ 
&+ \hbar \nu \left( \hat{a}_1^\dag \hat{a}_2^{\phantom \dagger} + \hat{a}_2^\dag \hat{a}_1^{\phantom \dagger} \right) + \hbar \left( G_1^{(a_1)} \hat{a}_1^\dag \hat{a}^{\phantom \dagger}_1 + G_1^{(a_2)} \hat{a}_2^\dag \hat{a}^{\phantom \dagger}_2 \right) \hat{x} \\ 
&+ \frac{\hbar}{2} \left( G_2^{(a_1)} \hat{a}_1^\dag \hat{a}^{\phantom \dagger}_1+ G_2^{(a_2)} \hat{a}_2^\dag \hat{a}^{\phantom \dagger}_2 \right) \hat{x}^2. 
\label{Htwooptexp}
\end{split}
\end{equation}
Choosing the optical modes to be degenerate (in the absence of coupling between them) such that $\omega_1 = \omega_2 = \omega_0$, we introduce the new basis with annihilation operators $\hat{a}_{\pm} = \left( \hat{a}_1 \pm \hat{a}_2 \right) / \sqrt{2}$, which describes the two supermodes that emerge due to the avoided crossing between the two original degenerate optical modes. The new Hamiltonian in this supermode basis can then be written as
\begin{equation}
\begin{split}
&\hat{H} = \hbar \omega_+ \hat{a}_+^\dag \hat{a}^{\phantom \dagger}_+ + \hbar \omega_- \hat{a}_-^\dag \hat{a}^{\phantom \dagger}_- + \hbar \omega_{\rm m} \hat{b}^\dag \hat{b} \\ 
&+ \hbar \left( \frac{G_1^{(a_1)} + G_1^{(a_2)}}{2} \right) \left( \hat{a}_+^\dag \hat{a}^{\phantom \dagger}_+ + \hat{a}_-^\dag \hat{a}^{\phantom \dagger}_- \right) \hat{x} \\
&+ \hbar \left( \frac{G_1^{(a_1)} - G_1^{(a_2)}}{2} \right) \left( \hat{a}_+^\dag \hat{a}^{\phantom \dagger}_- + \hat{a}_-^\dag \hat{a}^{\phantom \dagger}_+ \right) \hat{x} \\ 
&+ \hbar \left( \frac{G_2^{(a_1)} + G_2^{(a_2)}}{4} \right) \left( \hat{a}_+^\dag \hat{a}^{\phantom \dagger}_+ + \hat{a}_-^\dag \hat{a}^{\phantom \dagger}_- \right) \hat{x}^2 \\
&+ \hbar \left( \frac{G_2^{(a_1)} - G_2^{(a_2)}}{4} \right) \left( \hat{a}_+^\dag \hat{a}^{\phantom \dagger}_- + \hat{a}_-^\dag \hat{a}^{\phantom \dagger}_+ \right) \hat{x}^2, \\
\label{Hplusminus}
\end{split}
\end{equation}
where we now have the new supermode frequencies $\omega_\pm = \omega_0 \pm \nu$. The splitting between these two new supermodes is $\omega_+ - \omega_- = 2\nu$, such that each can be individually accessed if $\kappa_\pm < 2\nu$, with $\kappa_\pm$ being the linewidth of the mode corresponding to $\hat{a}_\pm$. Up to this point, we have not made any assumptions about the nature of the couplings in this system. In what follows, we will investigate how the Hamiltonian in Eq.~\eqref{Hplusminus} can be used to effectively describe an optomechanical MIM system, as well as a mechanical element quadratically-coupled to an optical mode via shared symmetries in a WGM optomechanical cavity.

\subsection{Membrane-in-the-Middle System}
\label{MIMsys}

In a conventional MIM optomechanical system, quadratic coupling arises due to the avoided crossing between the two hybridized optical supermodes mentioned above. Therefore, it is unnecessary to expand our optical frequencies to second order and we take $G_2^{(a_i)} = 0$ here. Furthermore, due to the geometry of MIM systems, as the mechanical elements is displaced, if the frequency of one optical mode increases, then the other mode's frequency will correspondingly decrease, leading to $G_1^{(a_1)} = -G_1^{(a_2)} = G_1$ \cite{jayich,paraiso}. As we shall see, this difference in sign between the linear coupling of the two optical modes is crucial for generating quadratic coupling in these systems, as well as enforcing the single-photon strong-coupling condition associated with using them for quantum nondemolition (QND) measurements of mechanical Fock states \cite{miao,ludwig,yanay}. Applying these conditions to the general two-mode optomechanical Hamiltonian in Eq.~\eqref{Hplusminus}, we obtain the Hamiltonian for a MIM system as \cite{paraiso,kalaee,miao,ludwig,yanay}
\begin{equation}
\begin{split}
&\hat{H}_{\rm MIM} = \hbar \omega_+ \hat{a}_+^\dag \hat{a}^{\phantom \dagger}_+ + \hbar \omega_- \hat{a}_-^\dag \hat{a}^{\phantom \dagger}_- + \hbar \omega_{\rm m} \hat{b}^\dag \hat{b} \\ 
&+ \hbar G_1 \left( \hat{a}_+^\dag \hat{a}^{\phantom \dagger}_- + \hat{a}_-^\dag \hat{a}^{\phantom \dagger}_+ \right) \hat{x} .
\label{HMIM}
\end{split}
\end{equation}
In this form, it is not obvious where the quadratic coupling arises in MIM systems. However, this system can be diagonalized, resulting in the Hamiltonian
\begin{equation}
\begin{split}
\hat{H}_{\rm MIM} = \hbar \omega'_{+} \hat{d}_+^\dag \hat{d}^{\phantom \dagger}_+ + \hbar \omega'_{-} \hat{d}_-^\dag \hat{d}^{\phantom \dagger}_- + \hbar \omega_{\rm m} \hat{b}^\dag \hat{b},
\label{Hdiag}
\end{split}
\end{equation}
with corresponding eigenfrequencies
\begin{equation}
\omega'_\pm = \omega_0 \pm \sqrt{\nu^2 + G_1^2 \hat{x}^2}.
\label{MIMfreq}
\end{equation}
In the limit where $\nu \gg \omega_{\rm m}$, $\hat{x}$ can be treated as a quasistatic variable \cite{miao,paraiso,ludwig,yanay}, allowing us to take $G_1 \hat{x} \ll \nu$. In this regime, the lowering operators of the diagonalized modes can be approximated as \cite{ludwig}
\begin{equation}
\begin{split}
\hat{d}_+ & \approx a_+ + \frac{G_1 \hat{x}}{2 \nu} \hat{a}_- , \nonumber \\
\hat{d}_- & \approx \frac{G_1 \hat{x}}{2 \nu} a_+ - a_-, \nonumber
\end{split}
\end{equation}
with the approximate frequencies
\begin{equation}
\omega'_\pm \approx \omega_0 \pm \left( \nu +  \frac{G_1^2}{2 \nu} \hat{x}^2 \right) = \omega_\pm \pm G_2' \hat{x}^2.
\label{MIMfreqapp}
\end{equation}
In this form, it is clear that these diagonalized mode frequencies exhibit a quadratic dependence on the position, with a coupling coefficient $G_2' = G_1^2/ 2 \nu$. Furthermore, the operators $\hat{d}_\pm$ are formed by a linear combination of the supermode operators $\hat{a}_\pm$, one of which is linearly coupled to the position variable $\hat{x}$. In this situation, even if we solely drive one of the supermodes, photons will tunnel to its counterpart and couple linearly to the mechanical resonator, causing its phononic Fock state to decohere. It is this process that leads to the condition of the single-photon strong-coupling regime ($g_1 \gg \kappa$) required to perform QND measurements of phonon states in MIM optomechanical systems \cite{miao,ludwig,yanay}.

\subsection{Whispering Gallery Mode System}
\label{WGMsys}

We now consider an optomechanical system whereby the motion of the mechanical element shifts the frequencies of both optical modes in the same direction. Such a system could be realized as a nanomechanical resonator side-coupled to an optical WGM cavity \cite{doolin}, with the two degenerate optical modes being the clockwise and counterclockwise propagating modes \cite{kippenberg}. In this case, the optomechanical coupling coefficients will be equal in both sign and magnitude, leading to $G_1^{(a_1)} = G_1^{(a_2)} = G_1$ and $G_2^{(a_1)} = G_2^{(a_2)} = G_2$. Note that we have opted to keep second-order terms in this analysis due to the fact that the avoided level crossing will no longer provide quadratic optomechanical coupling. Inserting these coefficients into the interaction Hamiltonian in Eq.~\eqref{Hplusminus}, we find
\begin{equation}
\begin{split}
&\hat{H}_{\rm WGM} = \hbar \omega_+ \hat{a}_+^\dag \hat{a}^{\phantom \dagger}_+ + \hbar \omega_- \hat{a}_-^\dag \hat{a}^{\phantom \dagger}_- + \hbar \omega_{\rm m} \hat{b}^\dag \hat{b} \\ &+ \hbar G_1 \left( \hat{a}_+^\dag \hat{a}^{\phantom \dagger}_+ + \hat{a}_-^\dag \hat{a}^{\phantom \dagger}_- \right) \hat{x} + \frac{\hbar G_2}{2} \left( \hat{a}_+^\dag \hat{a}^{\phantom \dagger}_+ + \hat{a}_-^\dag \hat{a}^{\phantom \dagger}_- \right) \hat{x}^2. 
\label{HWGM}
\end{split}
\end{equation}
For this system, we are thus left with a Hamiltonian that is already diagonalized. This has two very important consequences for this type of coupled system. First, the quadratic coupling that arose due to the avoided level crossing for the MIM system has vanished. However, there still exists quadratic coupling terms in our Hamiltonian as we have expanded the optical cavity resonance frequency to second order in mechanical position. Furthermore, as opposed to the MIM system, where the quadratic coupling is proportional to the square of the linear coupling, the second-order coupling coefficient here can be modified independently of the linear coupling by tuning the relative symmetry of the optical and mechanical modeshapes (see Appendix \ref{nondegen}). This leads us to the second important consequence of this system: since there is no linear mechanically-mediated coupling between the optical modes, QND measurements using this system are not constrained by the stringent single-photon strong-coupling regime. Instead, we obtain a limit on the linear coupling strength, $G_1$, with respect to the quadratic coupling, $G_2$, {\it i.e.}~Eq.~\eqref{linlim}.

\subsection{Mapping to a Single Optical Mode}
\label{singmap}

We conclude this section by noting that in the main text, we assumed a single optical mode, as opposed to the coupled two-mode system considered here. The effect of adding a second, undriven optical mode to the system can be included in the optically-induced transition rates given by Eqs.~\eqref{Gamn+1}-\eqref{Gamn-2}, \eqref{Gam1} and \eqref{Gam2} by taking $\bar{N} = \bar{N}_+ + \bar{N}_- = \bar{N}_1 + \bar{N}_2$, where $\bar{N}_i = \braket{\hat{a}_i^\dag \hat{a}^{\phantom \dagger}_i}$ is simply the average photon occupancy of the mode corresponding to $\hat{a}_i$. For such a system, if one drives the $\hat{a}_1$ mode (call it the clockwise mode) to a photon occupancy $\hat{N}_1$, then backscattering will cause the $\hat{a}_2$ mode (counterclockwise mode) to be populated to an occupancy \cite{wilson}
\begin{equation}
\bar{N}_2 = \frac{\nu^2}{\Delta^2 + \left( \kappa / 2 \right)^2} \bar{N}_1.
\label{N2occ}
\end{equation}
For the $\Delta = 0$ condition associated with the phase sensitive measurements discussed in Sec.~\ref{qndcond}, we then have $\bar{N}_2 = \left(2 \nu / \kappa \right)^2 \bar{N}_1$. For $2 \nu \ll \kappa$, $\bar{N}_2 \ll \bar{N}_1$, that is the counterclockwise mode is essentially unpopulated such that we can take $\bar{N} \approx \bar{N}_1$. Therefore, in this regime, we need only consider one mode (in this case the clockwise mode). We note that in this situation a small, but finite leakage of photons into the counterclockwise mode will not lead to accelerated decoherence associated with the MIM system \cite{miao} due to the fact that counterclockwise photons interact with the mechanics in the same way clockwise photons do.

For the case where $2 \nu \gg \kappa$, the situation is complicated by the fact that the clockwise and counterclockwise modes hybridize into the two individually accessible symmetric and antisymmetric modes corresponding to $\hat{a}_\pm$. Under these circumstances, the resonant probing condition required for phase sensitive measurements results in $\Delta = \pm \nu$. In either case this leads to $\bar{N}_2 \approx \bar{N}_1$, meaning that even though we are only driving the clockwise mode, strong backscattering ensures that in equilibrium both modes are equally populated. Again, due to the fact that photons in the clockwise and counterclockwise modes interact identically with the mechanics, this photon redistribution does nothing to affect the optically-induced transition rates. However, as half of the photons now reside in the unmonitored counterclockwise mode, the measurement rate is halved. Therefore, a single mode treatment is still valid in this regime, provided we account for this factor of two decrease in the measurement rate of the mechanical phonon number.

\section{Optomechanical Coupling using Non-Degenerate Perturbation Theory}
\label{nondegen}

Here we use the perturbative approach developed by Johnson {\it et al.}~\cite{johnson} to determine the first- and second-order optomechanical coupling coefficients for mechanical systems coupled to non-degenerate optical modes. In doing so, we will show that it is possible, in principle, to completely eliminate linear coupling in favor of quadratic coupling by exploiting the symmetry of an optomechanical system. 

We begin by considering a high-$Q$ optical mode with resonant angular frequency $\omega_i$, such that we can approximate the time-dependence of the mode's electric field as $\vec{E}_i(\vec{r},t) = e^{i \omega_i t} \vec{E}_i(\vec{r})$. Using Maxwell's equations for a source-free dielectric, as is relevant for the optomechanical systems considered here, one obtains the Helmholtz equation for the electric field as
\begin{equation}
\nabla^2 \ket{E_i} = -\frac{\omega_i^2 \epsilon_r(\vec{r})}{c^2} \ket{E_i},
\label{waveeq}
\end{equation}
where $\nabla^2$ is the Laplacian operator and the geometry of the resonator is specified by its spatially-varying relative permittivity profile $\epsilon_r(\vec{r})$. We have also chosen to follow the notation of Johnson {\it et al.}~by representing the electric field of the cavity mode using the Dirac braket state vectors $\ket{E_i} = \vec{E}_i(\vec{r})$, which have an inner product defined as 
\begin{equation}
\braket{ E_i | E_j} \equiv \int \vec{E}^*_i(\vec{r}) \cdot \vec{E}_j(\vec{r}) dV,
\end{equation}
where the integral is performed over the volume of the optomechanical system \cite{johnson}. With this definition, the optical modes of the cavity are orthogonal in the sense that $\bra{E_i} \epsilon_r \ket{E_j} = \bra{E_i} \epsilon_r \ket{E_i} \delta_{ij}$.

We now imagine introducing a small shift in the cavity's permittivity profile, resulting in $\epsilon_r(\vec{r}) \rightarrow \epsilon_r(\vec{r}) + \delta \epsilon_r(\vec{r})$. Treating the problem perturbatively, we determine the new electric fields $\ket{E'_i}$, and their corresponding frequencies $\omega'_i$, in this shifted geometry by expanding to second order as 
\begin{align}
\label{Esec}
\ket{E'_i} &= \ket{E_i^{(0)}} + \ket{E_i^{(1)}} + \ket{E_i^{(2)}}, \\
\label{omsec}
\omega'_i &= \omega_i^{(0)} + \omega_i^{(1)} + \omega_i^{(2)}. 
\end{align}
Here the superscript $(0)$ indicates the original unperturbed quantity, while the $(1)$ and $(2)$ indicate the first- and second-order corrections, proportional to $\delta \epsilon_r$ and $\left( \delta \epsilon_r \right)^2$, respectively. We note that these higher order corrections to the electric field are chosen to be orthogonal to the unperturbed field in the same sense as before such that $\bra{E_i^{(0)}} \epsilon_r \ket{E_i^{(n > 0)}} = 0$.

For perturbations that are optomechanical in nature, the shift of the dielectric profile will be induced due to the motion of a mechanical element. In this case, we can also expand the optical mode frequency to second order in a similar fashion to Eq.~\eqref{omtwoopt} as
\begin{equation}
\omega'_i = \omega_i + G_1 \Delta x + \frac{G_2}{2} (\Delta x)^2,
\label{omx}
\end{equation}
where $\Delta x$ is the resonator's displacement from equilibrium. Matching these terms with the ones found in Eq.~\eqref{omsec}, Eq.~\eqref{waveeq} can be solved order-by-order to find \cite{johnson,eichenfield,kaviani,rodriguez}
\begin{align}
\label{om0}
&\omega_i = \omega_i^{(0)}, \\
\label{om1}
& G_1 = \frac{\omega^{(1)}}{\Delta x} = - \frac{\omega^{(0)}}{2} \frac{\bra{E_i^{(0)}} \frac{d \epsilon_r}{dx} \ket{E_i^{(0)}}}{\bra{E_i^{(0)}} \epsilon_r \ket{E_i^{(0)}}}, \\
\label{om2}
&G_2 = \frac{\omega^{(2)}}{(\Delta x)^2} = \frac{3 G_1^2}{\omega_i} + \sum_{\omega_j \ne \omega_i} G_{ij},
\end{align}
where the sum is performed over all other optical cavity modes and
\begin{equation}
G_{ij} = \frac{\omega_i^3}{\omega_i^2-\omega_j^2} \frac{| \bra{E_j^{(0)}} \frac{d \epsilon_r}{dx} \ket{E_i^{(0)}} |^2}{\bra{E_i^{(0)}} \epsilon_r \ket{E_i^{(0)}} \bra{E_j^{(0)}} \epsilon_r \ket{E_j^{(0)}}}.
\end{equation}
Upon inspection of Eq.~\eqref{om1}, we see that linear coupling is proportional to the self-overlap of the optical mode, mediated by the change in relative permittivity with respect to the mechanical displacement, $x$. Meanwhile, in Eq.~\eqref{om2} the quadratic coupling exhibits both a self-overlap term, as well as a term depending on the cross-coupling between the original unperturbed modes and the spectrum of other non-degenerate cavity modes (the case of quadratic coupling in degenerate cavity modes was discussed in Appendix \ref{MIMWGM}). Therefore, linear optomechanical coupling will in principle be zero if the field self-overlap term vanishes, that is $\bra{E_i^{(0)}} \frac{d \epsilon_r}{dx} \ket{E_i^{(0)}} = 0$ \cite{kaviani}. Furthermore, in this situation the quadratic coupling is given by
\begin{equation}
G_2 = \sum_{\omega_j \ne \omega_i} G_{ij},
\label{G2noG1}
\end{equation}
with only the cross-coupling terms surviving. Therefore, provided these terms do not sum to zero, a non-zero quadratic coupling can be achieved in the absence of linear coupling \cite{kaviani}.

To better understand the physical conditions that lead to vanishing linear optomechanical coupling, we investigate the case where the optomechanical coupling is due to shifting the boundary conditions of the optical mode (as opposed to the photoelastic effect \cite{chan}), pertinent to the majority of optomechanical systems. For this situation, we find that \cite{johnson,eichenfield}
\begin{equation}
\bra{E_i} \frac{d \epsilon_r}{d x} \ket{E_j} = \int \left( \vec{q} (\vec{r}) \cdot \vec{u} \right) \left[ \Delta \epsilon \vec{E}_i^{\parallel *} \cdot \vec{E}^\parallel_j - \Delta \epsilon^{-1} \vec{D}_i^{\perp *} \cdot \vec{D}_j^\perp \right] dA,
\label{shiftbound}
\end{equation}
where $\vec{q}(\vec{r})$ is the mechanical modeshape function and $\vec{D} = \epsilon_0 \epsilon_r \vec{E}$ is the electric displacement field. The integral is performed over the surface of the unperturbed optical resonator as defined by its unit normal vector $\vec{u}$. We have also introduced the superscripts $\parallel$ and $\perp$ to denote the components of the associated fields parallel and perpendicular to the cavity surface.  Finally, $\Delta \epsilon = \epsilon_{\rm d} - \epsilon_{\rm s}$ and $\Delta \epsilon^{-1} = \epsilon_{\rm d}^{-1} - \epsilon_{\rm s}^{-1}$, where $\epsilon_{\rm d}$ and $\epsilon_{\rm s}$ are the relative permittivities of the optomechanical device's material and the surrounding medium, respectively.

The expression in Eq.~\eqref{shiftbound} will be zero if the integrand is an odd function with respect to the symmetry axes of the optical cavity. In practice, this can be realized by implementing an optical intensity profile that exhibits even symmetry, along with a mechanical modeshape (after dot product with the unit surface normal) that demonstrates odd symmetry \cite{kaviani}. This amounts to having an optical field which is unable to distinguish the direction of motion of the mechanics, that is the optical frequency shift is even with respect to mechanical displacement. Therefore, the first term in the cavity expansion (ignoring the zeroth-order term corresponding to the unperturbed cavity frequency) must be proportional to $(\Delta x)^2$, leading to quadratic optomechanical coupling.

\end{appendix}


\begin{thebibliography}{99}

\bibitem{joos}
E. Joos, H. D. Zeh, C. Kiefer, D. J. W. Giulini, J. Kupsch, and I.-O. Stamatescu, {\it Decoherence and the Appearance of a Classical World in Quantum Theory} (Springer-Verlag, Berlin, 2003).

\bibitem{schlosshauer}
M. A. Schlosshauer, {\it Decoherence and the Quantum-to-Classical Transition} (Springer-Verlag, Berlin, 2007).

\bibitem{ghirardi}
G. C. Ghirardi, A. Rimini, and T. Weber, Phys. Rev. D {\bf 34}, 470 (1986).

\bibitem{diosi}
L. Di\'osi, Phys. Rev. A {\bf 40}, 1165 (1989).

\bibitem{penrose}
R. Penrose, Gen. Relativ. Gravit. {\bf 28}, 581 (1996).

\bibitem{zurek}
W. H. Zurek, Rev. Mod. Phys. {\bf 75}, 715 (2003).

\bibitem{bose}
S. Bose, K. Jacobs, and P. L. Knight, Phys. Rev. A {\bf 59}, 3204 (1999).

\bibitem{marshall}
W. Marshall, C. Simon, R. Penrose, and D. Bouwmeester, Phys. Rev. Lett. {\bf 91}, 130401 (2003).

\bibitem{romeroisart}
O. Romero-Isart, A. C. Pflanzer, F. Blaser, R. Kaltenbaek, N. Kiesel, M. Aspelmeyer, and J. I. Cirac, Phys. Rev. Lett. {\bf 107}, 020405 (2011).

\bibitem{chen}
Y. Chen, J. Phys. B {\bf 46}, 104001 (2013).

\bibitem{aspelmeyer}
M. Aspelmeyer, T. J. Kippenberg, and F. Marquardt, Rev. Mod. Phys. {\bf 86}, 1391 (2014).

\bibitem{thompson}
J. D. Thompson, B. M. Zwickl, A. M. Jayich, F. Marquardt, S. M. Girvin, and J. G. E. Harris, Nature {\b 452}, 72 (2008).

\bibitem{clerk2}
A. A. Clerk, F. Marquardt, and J. G. E. Harris, Phys. Rev. Lett. {\bf 104} 213603 (2010).

\bibitem{gangat}
A. A. Gangat, T. M. Stace, and G. J. Milburn, New J. Phys. {\bf 13}, 043024 (2011).

\bibitem{vanner}
M. R. Vanner, Phys. Rev. X {\bf 1}, 021011 (2011).

\bibitem{borkje}
K. B{\o}rkje, Phys. Rev. A {\bf 90}, 023806 (2014).

\bibitem{riedinger}
R. Riedinger, S. Hong, R. A. Norte, J. A. Slater, J. Shang, A. G. Krause, V. Anant, M. Aspelmeyer, and S. Gr\"oblacher, Nature {\bf 530}, 313 (2016).

\bibitem{teufel}
J. D. Teufel, T. Donner, D. Li, J. W. Harlow, M. S. Allman, K. Cicak, A. J. Sirois, J. D. Whittaker, K. W. Lehnert, and R. W. Simmonds, Nature {\bf 475}, 359 (2011).

\bibitem{chan}
J. Chan, T. P. Mayer Alegre, A. H. Safavi-Naeini, J. T. Hill, A. Krause, S. Gr\"oblacher, M. Aspelmeyer, and O. Painter, Nature {\bf 478}, 89 (2011).

\bibitem{safavinaeini}
A. H. Safavi-Naeini, J. Chan, J. T. Hill, T. P. Mayer Alegre, A. Krause, and O. Painter, Phys. Rev. Lett. {\bf 108}, 033602 (2012).

\bibitem{safavinaeini2}
A. H. Safavi-Naeini, S. Gr\"oblacher, J. T. Hill, J. Chan, M. Aspelmeyer, and O. Painter, Nature {\bf 500}, 185 (2013).

\bibitem{palomaki}
T. A. Palomaki, J. D. Teufel, R. W. Simmonds, and K. W. Lehnert, Science {\bf 342}, 710 (2013).

\bibitem{suh}
J. Suh, A. J. Weinstein, C. U. Lei, E. E. Wollman, S. K. Steinke, P. Meystre, A. A. Clerk, and K. C. Schwab, Science {\bf 344}, 1262 (2014).

\bibitem{weinstein}
A. J. Weinstein, C. U. Lei, E. E. Wollman, J. Suh, A. Metelmann, A. A. Clerk, and K. C. Schwab, Phys. Rev. X {\bf 4}, 041003 (2014).

\bibitem{wollman}
E. E. Wollman, C. U. Lei, A. J. Weinstein, J. Suh, A. Kronwald, F. Marquardt, A. A. Clerk, and K. C. Schwab, Science {\bf 349}, 952 (2015).

\bibitem{pirkkalainen}
J.-M. Pirkkalainen, E. Damsk\"agg, M. Brandt, F. Massel, and M. A. Sillanp\"a\"a, Phys. Rev. Lett. {\bf 115}, 243601 (2015).

\bibitem{lecocq}
F. Lecocq, J. B. Clark, R. W. Simmonds, J. Aumentado, and J. D. Teufel, Phys. Rev. X {\bf 5}, 041037 (2015).

\bibitem{lei}
C. U. Lei, A. J. Weinstein, J. Suh, E. E. Wollman, A. Kronwald, F. Marquardt, A. A. Clerk, and K. C. Schwab, Phys. Rev. Lett. {\bf 117}, 100801 (2016).

\bibitem{wilson}
D. J. Wilson, V. Sudhir, N. Piro, R. Schilling, A. Ghadimi, and T. J. Kippenberg, Nature {\bf 524}, 325 (2015).

\bibitem{hong}
S. Hong, R. Riedinger, I. Marinkovi\'c, A. Wallucks, S. G. Hofer, R. A. Norte, M. Aspelmeyer, and S. Gr\"oblacher, Science {\bf 358}, 203 (2017).

\bibitem{reed}
A. P. Reed, K. H. Mayer, J. D. Teufel, L. D. Burkhart, W. Pfaff, M. Reagor, L. Sletten, X. Ma, R. J. Schoelkopf, E. Knill, and K. W. Lehnert, Nat. Phys. {\bf 13}, 1163 (2017).

\bibitem{riedinger2}
R. Riedinger, A. Wallucks, I. Marinkovi\'c, C. L\"oschnauer, M. Aspelmeyer, S. Hong, and S. Gr\"oblacher, Nature {\bf 556}, 473 (2018).

\bibitem{ockeleon-korppi}
C. F. Ockeleon-Korppi, E. Damsk\"agg, J.-M. Pirkkalainen, M. Asjad, A. A. Clerk, F. Massel, M. J. Wooley, and M. A. Sillanp\"a\"a Nature {\bf 556}, 478 (2018).

\bibitem{bergquist}
J. C. Bergquist, R. G. Hulet, W. M. Itano, and D. J. Wineland, Phys. Rev. Lett. {\bf 57}, 1699 (1986).

\bibitem{peil}
S. Peil and G. Gabrielse, Phys. Rev. Lett. {\bf 83}, 1287 (1999).

\bibitem{nogues}
G. Nogues, A. Rauschenbeutel, S. Osnaghi, M. Brune, J. M. Raimond, and S. Haroche, Nature {\bf 400}, 239 (1999).

\bibitem{gleyzes}
S. Gleyzes, S. Kuhr, C. Guerlin, J. Bernu, S. Del\'eglise, U. B. Hoff, M. Brune, J.-M. Raimond, and S. Haroche, Nature {\bf 446}, 297 (2007).

\bibitem{neumann}
P. Neumann, J. Beck, M. Steiner, F. Rempp, H. Fedder, P. R. Hemmer, J. Wrachtrup, and F. Jelezko, Science {\bf 329}, 542 (2010).

\bibitem{lupascu}
A. Lupa\c{s}cu, S. Saito, T. Picot, P. C. de Groot, C. J. P. M. Harmans, and J.E. Mooij, Nat. Phys. {\bf 3}, 119 (2007).

\bibitem{unruh}
W. G. Unruh, Phys. Rev. D {\bf 18}, 1764 (1978).

\bibitem{braginsky}
V. B. Braginsky and F. Y. Khalili, {\it Quantum Measurement}, edited by K. S. Thorne (Cambridge University Press, Cambridge, 1995).

\bibitem{braginsky2}
V. B. Bragisnky, Y. I. Vorontsov, and K. S. Thorne, Science {\bf 209}, 547 (1980).

\bibitem{clerk}
A. A. Clerk, M. H. Devoret, S. M. Girvin, F. Marquardt, and R. J. Schoelkopf, Rev. Mod. Phys. {\bf 82}, 1155 (2010).

\bibitem{jayich}
A. M. Jayich, J. C. Sankey, B. M. Zwickl, C. Yang, J. D. Thompson, S. M. Girvin, A. A. Clerk, F. Marquardt, and J. G. E. Harris, New. J. Phys. {\bf 10}, 095008 (2008).

\bibitem{sankey}
J. C. Sankey, C. Yang, B. M. Zwickl, A. M. Jayich, and J. G. E. Harris, Nat. Phys. {\bf 6}, 707 (2010).

\bibitem{purdy}
T. P. Purdy, D. W. C. Brooks, T. Botter, N. Brahms, Z.-Y. Ma, and D. M. Stamper-Kurn, Phys. Rev. Lett. {\bf 105}, 133602 (2010).

\bibitem{paraiso}
T. K. Para\"iso, M. Kalaee, L. Zang, H. Pfeifer, F. Marquardt, and O. Painter, Phys. Rev. X {\bf 5}, 041024 (2015).

\bibitem{kalaee}
M. Kalaee, T. K. Para\"iso, H. Pfeifer, and O. Painter, Opt. Express {\bf 24}, 21308 (2016).

\bibitem{miao}
H. Miao, S. Danilishin, T. Corbitt, and Y. Chen, Phys. Rev. Lett. {\bf 103}, 100402 (2009).

\bibitem{ludwig}
M. Ludwig, A. H. Safavi-Naeini, O. Painter, and F. Marquardt, Phys. Rev. Lett. {\bf 109}, 063601 (2012).

\bibitem{yanay}
Y. Yanay, J. C. Sankey, and A. A. Clerk, Phys. Rev. A {\bf 93}, 063809 (2016).

\bibitem{kaviani}
H. Kaviani, C. Healey, M. Wu, R. Ghobadi, A. Hryciw, and P. E. Barclay, Optica {\bf 2}, 271 (2015).

\bibitem{doolin}
C. Doolin, B. D. Hauer, P. H. Kim, A. J. R. MacDonald, H. Ramp, and J. P. Davis, Phys. Rev. A {\bf 89}, 053838 (2014).

\bibitem{hauerAOP1}
B. D. Hauer, C. Doolin, K. S. D. Beach, and J. P. Davis, Ann. Phys. {\bf 339}, 181 (2013).

\bibitem{carmichael}
H. J. Carmichael, {\it Statistical Methods in Quantum Optics 2: Non-Classical Fields} (Springer-Verlag, Berlin, 2008).

\bibitem{martin}
I. Martin and W. H. Zurek, Phys. Rev. Lett. {\it 98}, 120401 (2007).

\bibitem{marquardt}
F. Marquardt, J. P. Chen, A. A. Clerk, and S. M. Girvin, Phys. Rev. Lett. {\bf 99}, 093902 (2007).
 
\bibitem{santamore}
D. H. Santamore, A. C. Doherty, and M. C. Cross, Phys. Rev. B {\bf 70}, 144301 (2004).

\bibitem{gardiner}
C. W. Gardiner and P. Zoller, {\it Quantum Noise: A Handbook of Markovian and Non-Markovian Quantum Stochastic Methods with Applications to Quantum Optics} (Springer-Verlag, Berlin, 2004).

\bibitem{johansson}
J. R. Johansson, P. D. Nation, and F. Nori, Comp. Phys. Comm. {\bf 184}, 1234 (2013).

\bibitem{schliesser}
A. Schliesser, G. Anetsberger, R. Rivi\`ere, O.Arcizet, and T. J. Kippenberg, New. J. Phys. {\bf 10}, 095015 (2008).

\bibitem{rocheleau}
T. Rocheleau, T. Ndukum, C. Macklin, J. B. Hertzberg, A. A. Clerk, and K. C. Schwab, Nature {\bf 463}, 72 (2010).

\bibitem{riviere}
R. Rivi\`ere, S. Del\'eglise, S. Weis, E. Gavartin, O. Arcizet, A. Schliesser, and T. J. Kippenberg, Phys. Rev. A {\bf 83}, 063835 (2011).

\bibitem{riviere2}
R. Rivi\`ere, O. Arcizet, A. Schliesser, and T. J. Kippenberg, Rev. Sci. Intrum. {\bf 84}, 043108 (2013).

\bibitem{meenehan}
S. M. Meenehan, J. D. Cohen, S. Gr\"oblacher, J. T. Hill, A. H. Safavi-Naeini, M. Aspelmeyer, and O. Painter, Phys. Rev. A {\bf 90}, 011803(R) (2014).

\bibitem{macdonald}
A. J. R. MacDonald, G. G. Popowich, B. D. Hauer, P. H. Kim, A. Fredrick, X. Rojas, P. Doolin, and J.P. Davis, Rev. Sci. Instrum. {\bf 86}, 013107 (2015).

\bibitem{macdonald2}
A. J. R. MacDonald, B. D. Hauer, X. Rojas, P. H. Kim, G. G. Popowich, and J. P. Davis, Phys. Rev. A {\bf 93}, 013836 (2016).

\bibitem{misra}
B Misra and E. C. G. Sudarshan, J. Math. Phys. {\bf 18}, 756 (1977).

\bibitem{itano}
W. M. Itano, D. J. Heinzen, J. J. Bollinger, and D. J. Wineland, Phys. Rev. A {\bf 41}, 2295 (1990).

\bibitem{fischer}
M. C. Fischer, B. Guti\'errez-Medina, and M. G. Raizen, Phys. Rev. Lett. {\bf 87}, 040402 (2001).

\bibitem{patil}
Y. S. Patil, S. Chakram, and M. Vengalattore, Phys. Rev. Lett. {\bf 115}, 140402 (2015).

\bibitem{wang}
Y.-D. Wang and A. A. Clerk, Phys. Rev. Lett. {\bf 108}, 153603 (2012).

\bibitem{stannigel}
K. Stannigel, P. Komar, S. J. M. Habraken, S. D. Bennett, M. D. Lukin, P. Zoller, and P. Rabl, Phys. Rev. Lett. {\bf 109}, 013603 (2012).

\bibitem{hill}
J. T. Hill, A. H. Safavi-Naeini, J. Chan, and O. Painter, Nat. Comm. {\bf 3}, 1196 (2012).

\bibitem{bochmann}
J. Bochmann, A. Vainsencher, D. D. Awschalom, and A. N. Cleland, Nat. Phys. {\bf 9}, 712 (2013).

\bibitem{palomaki2}
T. A. Palomaki, J. W. Harlow, J. D. Teufel, R. W. Simmonds, and K. W. Lehnert, Nature {\bf 495}, 210 (2013).

\bibitem{andrews}
R. W. Andrews, R. W. Peterson, T. P. Purdy, K. Cicak, R. W. Simmonds, C. A. Regal and K. W. Lehnert, Nat. Phys. {\bf 10}, 321 (2014).

\bibitem{dellantonio}
L. Dellantonio, O. Kyriienko, F. Marquardt, and A. S. S{\o}renson, arXiv:1801.02438 (2018).

\bibitem{kippenberg}
T. J. Kippenberg, S. M. Spillane, and K. J. Vahala, Opt. Lett. {\bf 27}, 1669 (2002).

\bibitem{johnson}
S. G. Johnson, M. Ibanescu, M. A. Skorobogatiy, O. Weisberg, J. D. Joannopoulos, and Y. Fink, Phys. Rev. E {\bf 65}, 066611 (2002).

\bibitem{eichenfield}
M. Eichenfield, J. Chan, A. H. Safavi-Naeini, K. J. Vahala, and O. Painter, Opt. Express {\bf 17}, 20078 (2009).

\bibitem{rodriguez}
A. W. Rodriguez, A. P. McCauley, P.-C. Hui, D. Woolf, E. Iwase, F. Capasso, M. Loncar, and S. G. Johnson, Opt. Express {\bf 19}, 2225 (2011).

\end{thebibliography}
\end{document}